\documentclass{aa}
\usepackage{graphicx}  
                                          
\begin{document}                                                                                
   \title{Helical jets in blazars}

   \subtitle{Interpretation of the multifrequency long-term variability \\of 
             \object{AO 0235+16}}
                                                                                
   \author{L. Ostorero\inst{1}
          \and M. Villata\inst{2}
          \and C. M. Raiteri\inst{2}}
                                                                
   \offprints{l.ostorero@lsw.uni-heidelberg.de}
                                                                                
   \institute{Landessternwarte Heidelberg-K\"onigstuhl, K\"onigstuhl 12, D-69117
              Heidelberg, Germany\\
              \email{l.ostorero@lsw.uni-heidelberg.de}
    \and
             Istituto Nazionale di Astrofisica (INAF), Osservatorio Astronomico di 
             Torino, Via Osservatorio 20, 10025 Pino Torinese (TO), Italy \\
             \email{villata@to.astro.it, raiteri@to.astro.it}}

   \date{Received; accepted}
              
   \titlerunning{Helical jets in blazars. AO 0235+16}
     
   \authorrunning{L.\ Ostorero et al.}   
                                   
   \abstract{
        The long-term variability of the multiwavelength blazar emission can be interpreted 
        in terms of orientation variations of a helical, inhomogeneous, non-thermally emitting 
        jet, possibly caused by the orbital motion of the parent black hole in a binary system
        (Villata \& Raiteri \cite{vilrai99}).
        The helical-jet model is here applied to explain the quasi-periodic
        radio-optical light curves and the broad-band spectral energy distributions (SEDs) of 
        the BL Lac object \object{AO 0235+16}. Through a suitable choice of the model
        parameters, the helix rotation can well account for the periodicity of
        the main radio and optical outbursts and for the corresponding SED variability, while the 
        interspersed minor radio events could be interpreted as due either to some local 
        distortions of the helical structure or to other phenomena contributing to the source 
        emission. In particular, the probable existence of flow instabilities provides a viable  
        interpretation for the non-periodic features.

           }
   \maketitle

\section{Introduction}
Blazars, namely BL Lacertae objects and flat-spectrum radio quasars, belong to the class of 
active galactic nuclei (AGNs). 
According to the unified model of AGNs (see e.g. Urry \& Padovani \cite{urrpad95}), their emission is 
dominated by radiation produced in relativistic plasma jets, which are oriented at small angles with
respect to the line of sight; the radiation is therefore strongly beamed towards the observer.

Details of the jet origin are still unknown, but their formation is thought to be triggered by 
the presence of a supermassive black hole surrounded by an accretion disc, possibly 
belonging to a binary black hole system (BBHS) hidden in the centre of the AGN.
On one hand, this scenario is suggested by the elliptical morphology of BL Lac host galaxies, 
which are believed to originate from merging phenomena between spirals, naturally leading to the 
formation of massive binary black holes (Begelman et al. \cite{beg80}; Wilson \& Colbert \cite{wilcol95}; 
Huges \& Blandford \cite{hug03}).
On the other hand, several observational evidences such as bending, misalignment, wiggling and 
precession of jets (Begelman et al. \cite{beg80};
Camenzind \& Krockenberger \cite{camkro92};
Kaastra \& Roos \cite{kaaroo92};
Conway \& Wrobel \cite{conwro95}; 
Villata et al. \cite{vil98};
Villata \& Raiteri \cite{vilrai99} and references therein; 
Abraham \cite{abr00}; Britzen et al. \cite{bri01}),
often associated with knots superluminally moving along different-scale curved trajectories, 
as well as the periodicity discovered in the multiwavelength light curves of some of these 
sources, have in some cases been interpreted in terms of helical structures 
tightly related to BBHSs.

The helical-jet model proposed by Villata \& Raiteri (\cite{vilrai99}) describes how orientation 
variations of the different-frequency emitting parts of a curved jet with respect to
the line of sight, possibly caused by the orbital motion of the parent black hole in a BBHS, can be the 
cause of the observed changes in the spectral energy distribution (SED) of a blazar. 
The emission from the jet is non-thermal: the relativistic electron population is responsible for 
producing both the low-energy (from radio to UV--X-rays) synchrotron radiation and the high-energy (up to 
$\gamma$-rays) one through inverse-Compton (IC) scattering of the same synchrotron photons
(synchrotron-self-Compton process, hereafter SSC).

This model has already provided an interpretation for the huge X-ray spectral brightening 
of \object{Mkn 501} (Villata \& Raiteri \cite{vilrai99}), the low-energy SED variations of 
\object{S4 0954+65} (Raiteri et al. \cite{rai99}), and the changes in the overall SEDs of 
\object{S5 0716+71} (Ostorero et al. \cite{ost01}) and \object{ON 231} (Sobrito et al. \cite{sob01}).
The modelling of a constant helix rotation allows us now to simulate the long-term behaviour of the 
multifrequency light curves together with the SED time evolution.
If the data sampling is good enough and the source emission exhibits a periodic (or 
quasi-periodic) behaviour, one can apply the model and find the set of parameters which best 
reproduces the observed multifrequency data, thus providing both a test for the model and
a description of the corresponding jet features.

In this paper the model, fully described in Sect.\ 2, is applied to the case of the BL Lac object 
\object{AO 0235+16}, whose radio (and optical) light curves have recently revealed a
$\sim 5.7$ year quasi-periodicity (Raiteri et al. \cite{rai01}).
Model light curves and SEDs are compared with observational data in Sect.\ 3.
Possible interpretations of minor, non-periodic events are presented in Sect.\ 4. 
A final discussion is performed in Sect.\ 5.

\section{The helical-jet model}
The helical-jet model by Villata \& Raiteri (\cite{vilrai99}) foresees that,
in a BBHS scenario, the emitting jet is bent because of the orbital motion and by the 
interaction of its magnetized plasma with the surrounding medium, twisting in a rotating 
helical-shaped structure.

Since in Villata \& Raiteri (\cite{vilrai99}) only the synchrotron emission was taken into 
account, we recall here the main formulae, adding those expressions relevant to
the high-energy SSC emission (see also Raiteri et al. \cite{rai04}).
 
Let us consider a helical-shaped jet with the helix axis taken as $z$-axis of a 3-D 
reference frame. 
The helix has a pitch angle $\zeta$ and the line of sight forms an angle $\psi$ with  
 the $z$-axis.   
The length of the helical jet path can be expressed as
\begin{equation}
l(z)=\frac{z}{\cos \zeta}\,,\quad 0\le z \le 1\,, 
\end{equation}
and covers an azimuthal angle
\begin{equation}
\varphi(z)=az\,,
\end{equation}
where $a$ is the total angle covered by the helix.
The jet viewing angle at $z$ is
\begin{equation}                        \label{costeta}
\cos\theta(z)=\cos\psi\cos\zeta + \sin\psi\sin\zeta\cos(\phi-az)\,,
\end{equation}
where $\phi$ is the ``rotation angle'', i.e.\ the azimuthal difference between the line of sight and the
initial direction of the helical path.

The jet geometry described above is the apparent shape of the plasma flow;
the different light travel times of each jet region and the plasma velocity components due to the 
rotation of the helical structure have been taken into account. 

The jet is assumed to be inhomogeneous, similarly to that of the model of Ghisellini \& Maraschi 
(\cite{ghimar89}; see also Ghisellini et al. \cite{ghi85}; Maraschi et al. \cite{mar92}): 
each slice of it emits synchrotron radiation in a range of frequencies between 
$\nu'_{\mathrm{s,min}}$ and $\nu'_{\mathrm{s,max}}$ 
(primed quantities refer to the plasma rest reference frame), which are supposed to decrease as power 
laws as the distance $l$ from the apex increases {\footnote {In the Ghisellini \& Maraschi 
(\cite{ghimar89}) model, what is parametrized and described with power laws are the physical quantities 
of the jet, on which the emitted frequencies depend. Since we are interested in Doppler factor variation 
effects (whatever the physical details can be), we prefer a more direct, phenomenological approach, in 
order to lower the parameter number. If needed, one can go back to the physical characteristics.}}:
\begin{equation}                        \label{nusynmin}
\nu'_{\mathrm{s,min}}(l)=\nu'_{\mathrm{s,min}}(0)
\left(1+\frac{l}{l_1}\right)^{-c_1}\,,\quad c_1>0\,,
\end{equation}
\begin{equation}                        \label{nusynmax}
\nu'_{\mathrm{s,max}}(l)=\nu'_{\mathrm{s,max}}(0)
\left(1+\frac{l}{l_2}\right)^{-c_2}\,,\quad c_2>0\,,
\end{equation}
where $\nu'_{\mathrm{s,min}}(0)$ and $\nu'_{\mathrm{s,max}}(0)$ 
are the values at $l=z=0$, $l_1$ and $l_2$ are length scales. 
In the following we assume that 
$\nu'_\mathrm{s,min}(0)=\nu'_{\mathrm{s,max}}(0)\equiv \nu'_{\mathrm{s,0}}$. 
The jet emits also higher-energy radiation, via IC scattering of soft synchrotron photons off 
high-energy electrons (SSC process), in a range of frequencies
\begin{equation}                        \label{nucommin}
\nu'_\mathrm{c,min}(l)=\frac{4}{3}\gamma^{2}_\mathrm{min}
\nu'_{\mathrm{s,min}}(l)\,,
\end{equation}
\begin{equation}                        \label{nucommax}
\nu'_\mathrm{c,max}(l)=\frac{4}{3}\gamma^{2}_\mathrm{max}(l)
\nu'_{\mathrm{s,max}}(l)\,,
\end{equation}
where the electron Lorentz factor $\gamma$ is supposed to vary between $\gamma_{\mathrm{min}}=1$ and
\begin{equation}                        \label{gammamax}
\gamma_\mathrm{max}(l)=\gamma_\mathrm{0}
\left(1+\frac{l}{l_{\gamma}}\right)^{-c_{\mathrm{\gamma}}}\,,\quad c_{\gamma}\ge 0\,,
\end{equation}
being $\gamma_{\mathrm{0}}=\gamma_{\mathrm{max}}(0)$.

Actually, in order to take the Klein-Nishina effect into account, an upper limit is set 
to $\nu'_\mathrm{c,max}(l)$, in the form
\begin{equation} 
\nu'_\mathrm{c,max}(l)=\gamma_\mathrm{max}(l)\, {\rm min} \left \{ \frac{4}{3}
                       \gamma_\mathrm{max}(l) \nu'_{\mathrm{s,max}}(l),\  
                       \frac{m_{\mathrm{e}} c^{2}}{h} \right \} \,.
\end{equation}

The observed flux density at frequency $\nu$ is assumed to be 
\begin{equation}
F_\nu(\nu)\propto\delta^3\nu^{-\alpha_0}\,,
\end{equation}
where $\alpha_0$ is the power-law index of the local synchrotron spectrum
(fixed to be $\alpha_0=0.5$) and 
\begin{equation}                        \label{delta}
\delta =[\Gamma(1-\beta\cos\theta)]^{-1}
\end{equation}
is the beaming or Doppler factor, $\beta$ being the bulk velocity of the emitting plasma in units
of the speed of light, $\Gamma=(1-\beta^2)^{-1/2}$ the corresponding bulk Lorentz factor, and 
$\theta$ the angle between the velocity vector and the line of sight, which varies along 
the helical path as shown in Eq.\ (\ref{costeta}), hence causing a dependence of $\delta$ on the 
$z$ coordinate. 

Notice that the local viewing angle $\theta$ of the jet reaches a minimum when 
$\phi-az$ is zero or a multiple of $360^{\circ}$,
implying the maximization of the beaming effect for frequencies emitted from the corresponding 
jet segments. 

By introducing an intrinsic flux-density dependence on $l$, 
the observed synchrotron and IC flux densities at frequency $\nu$ coming from a jet slice of 
thickness $\mathrm{d}l$ can be expressed as 
\begin{equation}                        \label{deeffesyn}
\mathrm{d}F_{\nu,\mathrm{s}}(\nu)=K_{\mathrm{s}} \left(1+\frac{l}{l_{\mathrm{s}}}
\right)^{-c_{\mathrm{s}}}\delta^3(l)\nu^{-\alpha_0}\,\mathrm{d}l\,,\quad c_{\mathrm{s}}
\geq0\,,
\end{equation}
\begin{eqnarray}                                \label{deeffecom}
\mathrm{d}F_{\nu,\mathrm{c}}(\nu) & = & K_{\mathrm{c}}
\left(1+\frac{l}{l_{\mathrm{c}}}\right)^{-c_{\mathrm{c}}}
\delta^3(l)\nu^{-\alpha_0} \nonumber \\ 
~ & ~ & \times \, \ln \frac{\nu'_{\mathrm{s,max}}(l)}{\nu'_{\mathrm{s,min}}(l)}\,
\mathrm{d}l\,,\quad \hspace{1.3cm} c_{\mathrm{c}}\geq0\,,
\end{eqnarray}
where $l_{\mathrm{s}}$ and $l_{\mathrm{c}}$ are again length scales, and the constants 
$K_{\mathrm{s}}$ and $K_{\mathrm{c}}$ are considered independent of time, meaning that 
intrinsic variations of the flux are not allowed.
The respective total flux densities at frequency $\nu$ coming from the whole jet are 
obtained by integrating along all the portions contributing to that observed frequency: 
\begin{eqnarray}                                \label{effesyn}
F_{\nu,\mathrm{s}}(\nu) & = & K_{\mathrm{s}}\,\nu^{-\alpha_0}
\sum_i{\int_{\Delta z_{\mathrm{s},i}(\nu)}
{\left[1+\frac{l(z)}{l_\mathrm{s}}\right]^{-c_\mathrm{s}}}} \nonumber \\
~ & ~ & \times\, \delta^3(z)\frac{\mathrm{d}l}{\mathrm{d}z}\,\mathrm{d}z\,,
\end{eqnarray}
\begin{eqnarray}                                \label{effecom}
F_{\nu,\mathrm{c}}(\nu) & = & K_{\mathrm{c}}\,\nu^{-\alpha_{0}} 
 \sum_i{\int_{\Delta z_{\mathrm{c},i}(\nu)}
 {\left[1+\frac{l(z)}{l_\mathrm{c}}\right]^{-c_\mathrm{c}}}} \nonumber \\ 
~ & ~ & \times\, \delta^3(z) \frac{\mathrm{d}l}{\mathrm{d}z}\,
{\ln \frac{\nu'_{\mathrm{s,max}}(z)}{\nu'_{\mathrm{s,min}}(z)}}\,
\mathrm{d}z\,,
\end{eqnarray}
where ${\mathrm{d}l}/{\mathrm{d}z}=1/\cos \zeta$, and $\Delta z_{\mathrm{s,c};i}(\nu)$ is the 
$z$ interval corresponding to the {\textit i}-th jet segment emitting radiation at frequencies 
observed as $\nu$ along the line of sight, i.e.\ where the condition
\begin{equation} 
\delta(z)\nu'_\mathrm{s,c;min}(z)\leq\nu\leq\delta(z)\nu'_\mathrm{s,c;max}(z)
\end{equation}
is satisfied.
An explanation of the logarithmic term in Eqs.\ (\ref{deeffecom}) and (\ref{effecom}) can be found in 
Ghisellini \& Maraschi \cite{ghimar89}.

The total flux density at frequency $\nu$ is finally obtained by summing the synchrotron and IC contributions.

If the helix rotates, e.g.\ because of the orbital motion of the parent black hole in a BBHS, the 
$\phi$ angle, and hence $\theta$ [Eq.\ (\ref{costeta})], vary with time; this consequently makes 
the Doppler factor $\delta$ [Eq.\ (\ref{delta})] of each emitting portion $\mathrm{d}l$ 
of the jet time-dependent, thus causing a temporal evolution of the observed flux density 
$F_\nu(\nu)$, even if the jet emission has not intrinsically changed at all.

\section{Modelling the periodic long-term behaviour of AO 0235+16}   
The recently discovered periodic behaviour of radio (and optical) light curves of AO 0235+16
(Raiteri et al. \cite{rai01}) and the wealth of historical spectral coverage of this source
suggested us the advisability of applying the rotating helical-jet model, intrinsically periodic, 
to model the long-term trend of its multifrequency light curves and to describe the corresponding 
broad-band SED time evolution (preliminary results of this work have already been published
in Ostorero et al. \cite{ost03a} and Ostorero et al. \cite{ost03b}).

With this purpose, we assembled the historical light curves and composed the multi-epoch SED of 
AO ~0235+16, by collecting data at all available frequencies from the literature and homogenizing
them through suitable corrections for extinction effects.
In particular, optical and infrared magnitudes were converted into fluxes by using the 
absolute calibrations of Bessel (\cite{bess79}) for standard $UBVKF$ and Cousins' $RI$ filters,
Wamsteker (\cite{wam81}) for standard $JHL$ bands, and Allen (\cite{allen73}) for
standard $RIN$ filters; infrared-to-ultraviolet dereddening was computed by adopting the
extinction laws of Rieke \& Lebofsky (\cite{rieleb85}) and Cardelli et al. (\cite{car89}), assuming 
a Galactic extinction $A_B=0.341 \, \rm mag$ (from NED).
In the X-ray band, absorbed fluxes were corrected for total absorption fading by using the 
XSPEC procedure.

\subsection{The radio-optical light curves}   \label{LC}
  
\begin{figure*} [!hbtp]         \label{mw}
\centering
\caption[]{\textit{Left panels}: Observed optical (mJy) and radio (Jy) light curves of AO 0235+16.
\textit{a}: $R$ optical band ($R$ band data, and $B$ band data before $\rm JD=2449000$ converted 
into $R$ ones by adopting a mean colour index $<B-R>=1.65\pm 0.16$; see
Raiteri et al. \protect\cite{rai01} for details);
\textit{b,c,d}: 14.5, 8.0, 4.8 GHz radio bands (UMRAO);
\textit{e}: 1.4 GHz (dots; Green Bank telescope) and 1.5 GHz (crosses; NRAO) radio bands;
data have been drawn together to better define the light curve features;
\textit{f}: 880 MHz radio band (Green Bank telescope);
\textit{g}: 430 MHz radio band (Arecibo telescope). 
These light curves are based on data from:
Owen et al. (\protect\cite{owe78}),
Owen et al. (\protect\cite{owe80}),
Ulvestad et al. (\protect\cite{ulv81}),                         
Perley (\protect\cite{per82}),                                  
Briggs (\protect\cite{bri83}),                                  
Landau et al. (\protect\cite{lan83}),                           
Ulvestad et al. (\protect\cite{ulv83}),                
Altschuler et al. (\protect\cite{alt84}), 
Ulvestad \& Johnston (\cite{ulvjoh84}),               
Aller et al. (\protect\cite{all85}), 
Rudnick et al. (\protect\cite{rud85}),                
Brown et al. (\protect\cite{bro89}),                            
Salgado et al. (\protect\cite{salg99}),
Raiteri et al. (\protect\cite{rai01}) and references therein.
\textit{Right panels}: Helical-jet modelling of the typical multifrequency outbursts shown in the
left panels (see the text for a detailed description of the model parameters relevant to curves 
\textit{a$^\prime$}--\textit{g$^\prime$}): as the wavelength increases, the outbursts appear to be 
more and more delayed and to have smaller amplitude and longer duration, in agreement with 
observations.
Grey (yellow in the electronic version) strips highlight here the time interval between the peak
of the modelled optical outburst and the maximum of the associated lowest-energy radio event.
The same strips have been drawn on the observed light curves of the left panels, in correspondence to the
periodic outbursts, even if in more than one case some counterparts are poorly sampled 
or missing at all.}
\end{figure*}

The most time-extended and best-sampled light curves of the source are the optical ones (see 
Fig.\ 1\textit{a}), and the radio ones at 14.5 GHz (Fig.\ 1\textit{b}), 8.0 GHz 
(Fig.\ 1\textit{c}) and 4.8 GHz (Fig.\ 1\textit{d}) from the University of Michigan Radio 
Astronomy Observatory (UMRAO). 
All these curves exhibit strong variability: many large-amplitude outbursts lasting from several
months to a few years witness an intense source activity. 
From the long-term time analysis of the best-sampled light curve at 8.0 GHz, Raiteri et al. 
(\cite{rai01}) found the signature of a quasi-periodic behaviour, also recognizable in the other
radio bands: the major outbursts repeat quasi-regularly every $\sim 5.7$ years, many of them accompanied 
by simultaneous or slightly preceding optical ones.
A visual inspection of the curves in the left panels of Fig.\ 1 reveals that the optical events 
are however sharper, more pronounced, and sometimes more frequent than the radio ones.
Indeed, Takalo et al. (\cite{tak98}) invoked the existence of two different optical emission 
mechanisms: one also responsible for the radio events (likely synchrotron processes in the 
relativistic jet), and the other not (probable microlensing effect of a foreground galaxy on 
radiation coming from either the accretion disc or other jet regions).
Moreover, the radio outbursts have decreasing amplitude with increasing wavelength, and flux 
variations at the lower frequencies lag those at the higher ones (see the cross-correlation 
analysis performed by Raiteri et al. \cite{rai01}); other minor outbursts are present in the radio 
light curves beside the periodic events.
Finally, all outbursts and flaring events seem to be superimposed to a slowly decreasing ``base 
level'' flux (see Sect.\ 3.2 for details).

Confirmations of these behaviours come from the lower-energy radio light curves, notwithstanding 
their poor sampling.
The delayed counterparts of both the 1982 periodic radio outburst and the 1980--81 non-periodic event
are likely represented by the couples of peaking structures sequentially appearing in the
curves at 1.4 GHz (Fig.\ 1\textit{e}; dots) and 880 MHz (Fig.\ 1\textit{f}) from the 
Green Bank telescope, at 1.5 GHz (Fig.\ 1\textit{e}; crosses) from the NRAO, and at 430 MHz 
(Fig.\ 1\textit{g}) from the Arecibo telescope.
Furthermore, the outburst peak flux seems to decrease with increasing wavelength more rapidly in 
the case of the periodic events than in the case of the non-periodic ones. Thus, while at high 
radio frequencies the periodic outbursts dominate the non-periodic ones, at low energies the vice 
versa seems to occur; this behaviour implies different radio broad-band
spectral indices for the periodic and non-periodic components. 
Moreover, the light curve at 430 MHz (Fig.\ 1\textit{g}) also exhibits 
the presence of a decreasing ``base level'' flux, on which the peaks corresponding to the 
periodic trend fade almost completely away. Notice also that the ``base level'' flux seems not 
to decrease with decreasing radio frequency (a similar behaviour was recognized in the BL Lac 
object S5 0716+71; see Raiteri et al. \cite{rai03}).

As anticipated in the previous section, the rotation of an inhomogeneous helical jet 
may account for the periodic behaviour of multiwavelength light curves. 
As the helix rotates, different-frequency emitting portions of the jet approach the line of sight, 
giving rise to outbursts observed at lower and lower energies as long as the rotation angle 
($\phi$) increases: this phenomenon comes from both the jet inhomogeneity and the beaming effect, 
as it will be discussed in the following.

The radio-optical light curve modelling was performed by assuming a period of 2069 days (see 
Raiteri et al. \cite{rai01}) for the constant helix rotation, and by fixing
$\phi=0$ at $\rm JD=2451032.6341$, corresponding to the 1998 outburst observed peak ($R$ band).

The strongest constraints to be taken into account in setting the model parameters are the 
observed outburst amplitudes, their durations, and the shortness of the delay of the radio events
with respect to the optical ones, which strongly depend on the inhomogeneity of the first part of the jet.

This inhomogeneity was modelled to meet the observational constraints by replacing the general single 
power-law trends for the emitted frequencies of the helical-jet model [Eqs. (\ref{nusynmin}) and 
(\ref{nusynmax})] with more complex laws.

In particular, the power law describing the minimum synchrotron frequency emitted along the jet path,
$\nu'_{\mathrm{s,min}}(l)$, was substituted by a combination of power laws for 
$l \leq l_{\mathrm{break}}$:

\begin{eqnarray}                        \label{nusynminlc}
\nu'_{\mathrm{s,min}}(l) & = & \nu'_{\mathrm{s,0,I}}
                \left(1+\frac{l}{l_{\mathrm{1,I}}}\right)^{-c_{\mathrm{1,I}}}+  \nonumber\\
& & \nu'_{\mathrm{s,0,II}}\left(1+\frac{l}{l_{\mathrm{1,II}}}\right)^{-c_{\mathrm{1,II}}}\,,
\quad c_{\mathrm{1,I}}\,,c_{\mathrm{1,II}}>0\,,
\end{eqnarray}
and with the tangent to $\log \nu'_{\mathrm{s,min}}(l)$ in $l=l_{\mathrm{break}}$, for 
$l~>~l_{\mathrm{break}}$,
while the $\nu'_{\mathrm{s,max}}(l)$ relation is simply a linear combination of power laws like 
that of Eq.\ (\ref{nusynmax}):
\begin{eqnarray}                        \label{nusynmaxlc}
\nu'_{\mathrm{s,max}}(l) & = & \nu'_{\mathrm{s,0,I}}
                \left(1+\frac{l}{l_{\mathrm{2,I}}}\right)^{-c_{\mathrm{2,I}}}+  \nonumber\\
& & \nu'_{\mathrm{s,0,II}}\left(1+\frac{l}{l_{\mathrm{2,II}}}\right)^{-c_{\mathrm{2,II}}}\,,
\quad c_{\mathrm{2,I}}\,,c_{\mathrm{2,II}}>0\,.
\end{eqnarray}
Both relations are represented by the continuous (blue) curves of Fig.\ 2 (the
parameter values are given in Table 1, Col.\ \textit{A}).

\begin{figure} [!hbtp]          \label{nusyn_fad}
\centering
\includegraphics[width=6cm,height=9.5cm,angle=90]{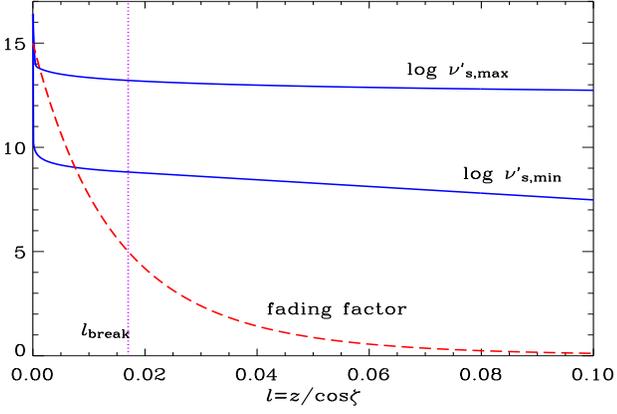}
\caption[]{
Solid (blue) curves represent the minimum and maximum synchrotron frequencies emitted by the 
modelled jet of AO ~0235+16, as a function of the helical path coordinate $l$.
Note the steepness of the $\nu'_{\mathrm{s,min}}(l)$ law in the first jet slice and its 
slowlier decreasing trend right after the apex region. 
The vertical, dotted (purple) line represents the point $l_{\mathrm{break}}=0.017$ where the law 
has been replaced by a linear decreasing trend. 
The law parameters are visible in Col.\ \textit{A} of Table 1. 
The dashed (red) curve shows the decrease, along the jet, of the intrinsic fading factor: it 
reduces dramatically the emission well before the first tenth of the path, making the jet
region far from the apex marginally contributing to the whole emission.
Only the portion of the path between $l=0$ and $l=0.1$ has been indeed represented here.
For graphic convenience, the fading factor has been normalized to 15.}
\end{figure}

The minimum emitted frequency law is very steep at first, implying the emission of frequencies
down to radio bands from the immediate vicinity of the apex of the emitting jet. 
This allows the observed optical and radio outbursts to follow one another with short time delays, 
as the contiguous jet slices mainly producing them become more aligned with the line of sight during the 
helix rotation.  
Then, the linear decrease of $\log \nu'_{\mathrm{s,min}}$ makes the emission of the observed 
very soft radio photons possible far from the jet apex, before the intrinsic flux-density 
dependence on $l$, represented by the ``fading factor'' 
$(1+l/l_{\mathrm{s}})^{-c_{\mathrm{s}}}$ [see Eq.\ (11)],
strongly reduces the jet emission (see the dashed, red curve of Fig.\ 2).

The results of the helical-jet modelling of the light curves of Fig.\ 1 (\textit{a}--\textit{g})
are displayed in the right panels of Fig.\ 1 (\textit{a$^\prime$}--\textit{g$^\prime$}).
Since the model light curves are exactly periodic, only snapshots including outbursts are 
shown for a visual comparison between the modelled and observed features.
Grey (yellow in the electronic version) stripes, defined as the time interval 
($\sim 242$ days) between the optical theoretical peak and the lowest-frequency radio one, guide 
the eye through the periodic outbursts in different energy bands.
Besides the periodic recurrence, also the typical shape of the main observed optical and radio 
outbursts, the decrease of the peak flux with decreasing radio frequency, and the optical-to-radio 
flux ratio are well described by the model.
The quasi-simultaneity of optical and radio outbursts also is well reproduced: radio delays start
from about a couple of months in the case of the shortest wavelengths, consistently with the 
results of the cross-correlation analysis performed by Raiteri et al. (\cite{rai01}), 
and gradually increase up to $\sim 8$ months, as one moves towards the lowest frequencies.

A complete summary of the parameters describing the geometrical and physical properties of 
the modelled AO 0235+16 jet is given in Table 1: both sets \textit{A} and \textit{B}
(see Sect.\ 3.2 for details) give rise to the same radio-to-optical light curves.
The emerging picture for the jet of AO 0235+16 is a  rotating helical-shaped structure with a 
pitch angle $\zeta=20^{\circ}$ and the axis oriented at an angle $\psi = 15^{\circ}$ with 
respect to the line of sight, implying for each jet slice a maximum alignment of $5^{\circ}$. 
The plasma flows along the helix with a bulk Lorentz factor $\Gamma =10$, and emits 
synchrotron and IC radiation, all photons mainly coming from a jet region close to the apex.

In general, the parameter values reported in Table 1 represent almost the
unique set of parameters which can strictly account for the multifrequency
light-curve (and SED, see below) behaviour, due to the strong constraints
imposed by the exceptional dataset.
However, some parameters, namely the jet geometry, $\Gamma$, and the IC
parameters, are less constrained due to two independent reasons.
For the former ones, what is strictly constrained by the data are the Doppler
factor variations, so that other combinations of $\Gamma$ and angles can work
as well. On the other hand, the high-energy part of the SED is poorly sampled
(see next subsection) and does not allow one to identify a unique choice of parameters.
In conclusion, the phenomenological parameters (and the corresponding physical
quantities, see footnote 1) we have here derived to match the data can
represent the ``most reasonable'' jet features, but not the sole solution.

\subsection{The SED time evolution}

\begin{figure*} [!hbtp]                 \label{seda}    
\centering
\includegraphics[width=16cm]{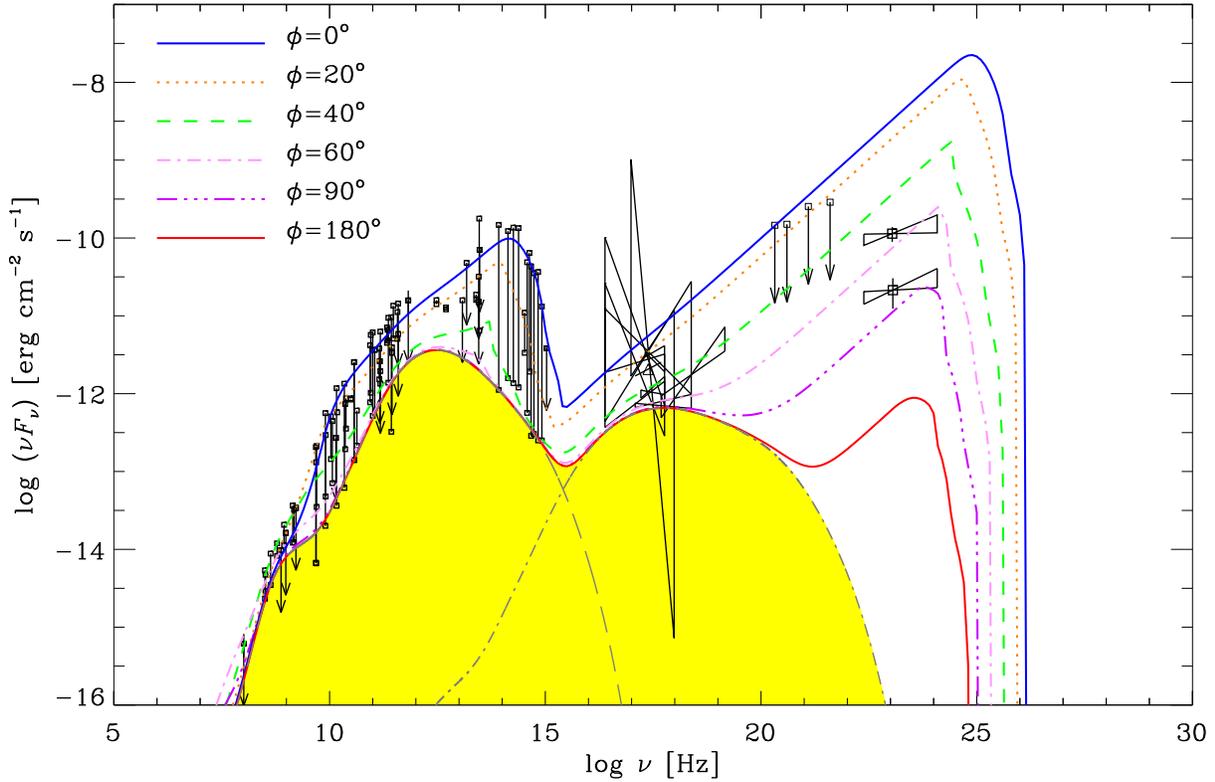}
\caption[]{
Multi-epoch SED of AO 0235+16. 
At radio, infrared, optical and ultraviolet frequencies, squares represent historical variation 
ranges when joined by vertical bars and single sporadic data otherwise; data are from:
Condon \& Jauncey (\protect\cite{conjau74}),
Ryle et al. (\protect\cite{ryl75}),                     
Altschuler \& Wardle (\protect\cite{altwar76}),
Rieke et al. (\protect\cite{rie76}),
O'Dell et al. (\protect\cite{ode77}),
Rieke et al. (\protect\cite{rie77}),
O'Dell et al. (\protect\cite{ode78a}),
O'Dell et al. (\protect\cite{ode78b}),
Owen et al. (\protect\cite{owe78}),                     
Condon et al. (\protect\cite{con79}),
Epstein et al. (\protect\cite{eps80}),
Landau et al. (\protect\cite{lan80}),
Owen et al. (\protect\cite{owe80}),
Weiler \& Johnston (\protect\cite{weijoh80}),
B\aa \aa th et al. (\protect\cite{baa81}),
Jones et al. (\protect\cite{jon81}),                    
Altschuler (\protect\cite{alt82}),
Ennis et al. (\protect\cite{enn82}),
Impey et al. (\protect\cite{imp82}),
Altschuler (\protect\cite{alt83}), 
Briggs (\protect\cite{bri83}),          
Landau et al. (\protect\cite{lan83}),
Seielstad et al. (\protect\cite{sei83}),
Sitko et al. (\protect\cite{sit83}),
Ulvestad et al. (\protect\cite{ulv83}),               
Altschuler et al. (\protect\cite{alt84}),               
Cotton et al. (\protect\cite{cot84}),
Cruz-Gonzales \& Hucra (\protect\cite{cruhuc84}),
Gear et al. (\protect\cite{gea84}),
Holmes et al. (\protect\cite{hol84}),
Impey et al. (\protect\cite{imp84}),
Ulvestad \& Johnston (\protect\cite{ulvjoh84}),
Gear et al. (\protect\cite{gea85}),
Moles et al. (\protect\cite{mol85}),
Rudnick et al. (\protect\cite{rud85}),          
Sitko et al. (\protect\cite{sit85}),
Brindle et al. (\protect\cite{bri86}),
Gear et al. (\protect\cite{gea86}),
Jones et al. (\protect\cite{jon86}),                    
Edelson (\protect\cite{ede87}),
Haddock \& Aller (\protect\cite{hadall87}),
Salonen et al. (\protect\cite{salo87}),
Smith et al. (\protect\cite{smi87}),
Fugmann \& Meisenheimer (\protect\cite{fugmei88}),
Impey \& Neugebauer (\protect\cite{impneu88}),
Steppe et al. (\protect\cite{ste88}),           
Wardle \& Roberts (\protect\cite{warrob88}),
Webb et al. (\protect\cite{web88}),
Brown et al. (\protect\cite{bro89}),                            
Vetukhnovskaya (\protect\cite{vet89}),
Impey \& Tapia (\protect\cite{imptap90}),
Mead et al. (\protect\cite{mea90}),
Simonetti \& Cordes (\protect\cite{simcor90}),
Gregory \& Condon (\protect\cite{grecon91}),
Sitko \& Sitko (\protect\cite{sitsit91}),                               
Edelson et al. (\protect\cite{ede92}),
Quirrenbach et al. (\protect\cite{qui92}),              
Takalo et al. (\protect\cite{tak92}),
Ter\"asranta et al. (\protect\cite{ter92}),             
Gear (\protect\cite{gea93}),
Krichbaum et al. (\protect\cite{kri93}),                
Steppe et al. (\protect\cite{ste93}),           
Bloom et al. (\protect\cite{blo94}),
Gear et al. (\protect\cite{gea94}),
Ghosh et al. (\protect\cite{gho94}),
Lichtfield et al. (\protect\cite{lic94}),
Tornikoski et al. (\protect\cite{tor96}),
Cotton et al. (\protect\cite{she97}),
Reuter et al. (\protect\cite{reu97}),
Shen et al. (\protect\cite{she97}),
Nartallo et al. (\protect\cite{nar98}),
Takalo et al. (\protect\cite{tak98}),
Salgado et al. (\protect\cite{salg99}),
Raiteri et al. (\protect\cite{rai01}) and references therein.
At high-energies, all data have been drawn: X-ray data (Einstein, EXOSAT, ROSAT, ASCA, RXTE)
are from
Worrall \& Wilkes (\protect\cite{worwil90}),
Ghosh \& Soundararajaperumal (\protect\cite{ghosou95}),
Madejski et al. (\protect\cite{mad96}), 
Comastri et al. (\protect\cite{com97}), 
Webb et al. (\protect\cite{web00}),
and the TARTARUS database;
$\gamma$-ray data (COMPTEL and EGRET) are from Sch\"onfelder et al. (\protect\cite{scho00}), 
Hartman et al. (\protect\cite{har99}), and Raiteri et al. (\protect\cite {rai01}).
The shaded (yellow) area shows the ``base level'' flux (synchrotron component: dashed line; 
IC component: dashed-dotted line).
Different-style (and colour) lines represent the SED evolution modelled with the helix 
rotation, i.e.\ with different values of the rotation angle $\phi$ (see the legend in the figure), 
under the hypothesis that the whole X-ray emission comes from IC scattering of
synchrotron photons off jet electrons. 
Model parameters are shown in Col.\ \textit{A} of Table 1.}
\end{figure*}

\begin{figure*} [!hbtp]                 \label{sedb}    
\centering
\includegraphics[width=16cm]{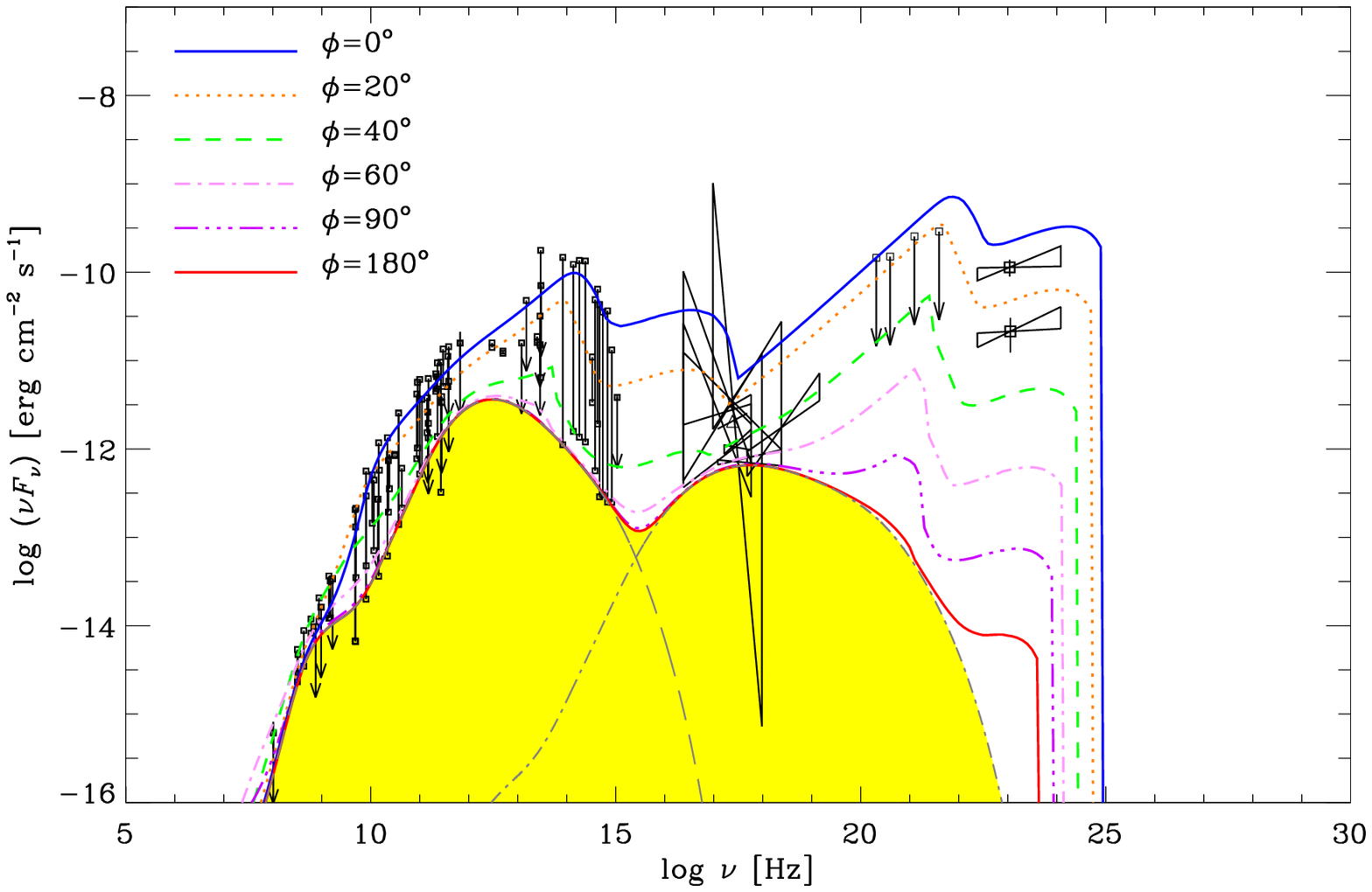}
\caption[]{
Multi-epoch SED of AO 0235+16. Historical data and ``base level'' flux are the same of the SED 
in Fig.\ 3. 
Helical-jet model fits have been here performed by supposing the existence of a synchrotron 
contribution to the soft X-ray emission, while the hard X-rays have still an IC origin.
Model parameters are shown in Col.\ \textit{B} of Table 1.}
\end{figure*}

The BL Lac object AO 0235+16 has been monitored in a wide range of frequencies since 1970's.
Radio observations were performed by many antennas, starting from about 100 MHz up to 
300 GHz; infrared space and ground data are available from submillimetric (far-IR) down to 
micron wavelengths (near-IR); optical $UBVRI$ bands have been intensively monitored by many 
telescopes all around the world, while only one ultraviolet upper limit by IUE is available.
AO 0235+16 was also detected in the X-ray band by several space observatories
operating within the range 0.5--60 keV (Einstein, EXOSAT, ROSAT, ASCA, RXTE)
starting from 1978.
The $\gamma$-ray emission was investigated through the instruments onboard the Compton 
Gamma-Ray Observatory (CGRO):
COMPTEL probed the low--medium energy $\gamma$-ray range of the source spectrum during CGRO 
Cycle 4 (1994--1995), and only set upper limits to the flux in the interval 0.75--30 MeV;
EGRET explored the higher-energy $\gamma$-ray radiation (30 MeV--20 GeV) with six pointings between 
1992 and 1997, and high-confidentially identified AO 0235+16 as a powerful $\gamma$-ray emitter,
providing two detections and four upper limits.

Figs.\ 3 and 4 show the source multi-epoch broad-band SEDs, composed by collecting literature data at
all available wavelengths: the historical variation ranges of the emission from radio to 
optical bands are represented together with all the X-ray and $\gamma$-ray data.
The low-energy part of the SED exhibits a variability which becomes less pronounced moving
from optical bands towards soft radio frequencies, as it is also evident from the previously 
discussed light curves of Fig.\ 1.
A variability range and a clear broad-band behaviour cannot instead be easily identified for the 
high-energy spectral component, because of the poor sampling of the X-ray and $\gamma$-ray light 
curves, and the uncertainties in data analysis, as detailed in the following.

In the X-ray energy domain, the source reveals a remarkable variability: 
the 1 keV flux spans less than one order of magnitude, but the spectral slope shows an 
intriguing alternance of steepening and hardening in the energy range 0.5--10 keV, the  
spectral index $\alpha$ ($F_{\nu}\propto \nu ^{-\alpha}$) ranging from 0.41 to 2.25.
In particular, the spectrum seems to be steeper when the source is in a brighter state, as already 
noticed by Madejski et al. (\cite{mad96}), who analysed the spectral changes of the 
source from the 1993 ROSAT detections to the 1994 ASCA ones:
such a behaviour, in disagreement with a simple synchrotron cooling model for the X-ray emission, 
could reveal that in AO 0235+16 two different spectral components overlap in the X-ray band, the 
drop in the intensity of the softer one 
uncovering the harder one. 
However, the softer component, which could be the tail of the synchrotron spectrum, is not so 
clearly identified, because of the large uncertainties in the X-ray spectral indices (see the 
X-ray butterflies in Figs.\ 3 and 4).
The main problem in the data analysis concerns indeed the estimate of the soft X-ray absorption. 
A model with a power-law continuum and absorption usually fits well the spectrum of AO 0235+16, and
the best fit is obtained by letting the hydrogen column density parameter $N_{\mathrm{H}}$ free; 
in this case, however, $N_{\mathrm{H}}$ is larger than the standard value. Alternatively,
one can keep $N_{\mathrm{H}}$ fixed, obtaining smaller uncertainties on the spectral indices but 
a worse fit.
From spectral studies, Madejski et al. (\cite {mad96}) concluded that the excess absorption effect is 
likely due to an intervening galaxy located at $z=0.524$, and performed a more
accurate estimate of the spectral indices by introducing the absorber at $z=0.524$ in the analysis.
However, the ambiguity on the X-ray spectral behaviour of the source is not completely solved,  
and the debate on the origin of X-rays in AO 0235+16 remains still open.

At $\gamma$-ray energies, the two clear detections by EGRET in 1994 and 1998 are not enough to 
constrain the source variability range. 
Only the 1994 signal is sufficiently strong to produce a useful spectrum, with a photon index 
$\gamma = 1.88 \pm 0.11$ (Mukherjee et al. \cite{muk97}; Hartman, private communication), even if 
a similar index $\gamma = 1.85 \pm 0.12$ was derived from the sum of the observations
of all viewing periods from Cycle 1 up to Cycle 4 (P1234; see Hartman et al. \cite{har99}).

As already noticed for the light curves of Fig.\ 1, the multiwavelength flux changes 
appear to be superimposed to a possible slowly variable ``base level'' emission, the deepest 
minimum of which is represented by the shaded (yellow) area of Figs.\ 3 and 4: 
the low-energy part of it was determined by extracting the most reliable historical minima from 
well-sampled radio, infrared, and optical light curves, and by fitting the corresponding SED 
with a cubic spline interpolation, while the high-energy part was analytically modelled 
according to the SSC theory, under the assumption that electron energies vary from 
$\gamma=1$ to $\gamma=10^{5}$. 

The long-term behaviour of the observed SED variability upon the ``base level'' was 
modelled as the result of the rotation of the steadily-emitting helical-shaped jet previously 
described.
In particular, the low-energy SED evolution is consistent with the radio-optical light 
curve modelling, while for the high-energy part two independent cases were 
considered to take the observation constraints into account:
the soft X-ray emission is, in the first case, IC radiation, while in the second case it also has a
synchrotron contribution, which is found to dominate for small ($\la 20^{\circ}$) rotation angles.
Different-style (and colour) curves in Figs.\ 3 and 4 show the model predictions for the SED time 
evolution in the two different hypothesis, as described in more details below.

\begin{table}[!hbtp]           
\caption{Helical-jet model input parameters relevant to light curve and SED evolution modelling of 
AO 0235+16.
Col.\ \textit{A}: light curves of Fig.\ 1 (right panels) and Fig.\ 5\textit{a}, and SED time evolution
of Fig.\ 3. 
Col.\ \textit{B}: SED time evolution of Fig.\ 4.
Values differing in the two models are written in boldface.}

\begin{tabular}{lcc}    
\hline
\textit{Jet geometry}                                   & \textit{A}      & \textit{B}   \\
\hline 
$\zeta$                                                         & 20$^{\circ}$    & 20$^{\circ}$ \\   
$\psi$                                                          & 15$^{\circ}$    & 15$^{\circ}$  \\
$a$                                                             & 360$^{\circ}$   & 360$^{\circ}$ \\ 
\hline
\textit{Jet physics}                                            & \textit{A}      & \textit{B} \\
\hline
$\Gamma$                                                        & 10            & 10  \\
$\gamma_{\mathrm{min}}$                                         & 1             & 1    \\
$\gamma_{\mathrm{0}}$                           & \boldmath{$1.995\times 10^{5}$}  & \boldmath{$6.310\times 10^{3}$}\\$c_{\gamma}$                                                    & 0             & 0             \\   
\hline
\textit{Synchrotron radiation}                          & \textit{A} & \textit{B} \\
\hline
$\alpha_{0}$                                                    & 0.5   & 0.5                   \\
$\nu '_{\mathrm{s,0,I}}$                                        & $10^{14}$     & $10^{14}$     \\
$\nu '_{\mathrm{s,0,II}}$                       & \boldmath{$1.259\times 10^{14}$} & \boldmath{$2.512\times 10^{16}$} \\         
$l_{\mathrm{break}}$                                            & 0.017         & 0.017                 \\
$l_{\mathrm{1,I}}=l_{\mathrm{1,II}}$                            & $10^{-10}$      & $10^{-10}$  \\
$l_{\mathrm{2,I}}=l_{\mathrm{2,II}}$                            & $10^{-3}$     & $10^{-3}$           \\
$c_{\mathrm{1,I}}=c_{\mathrm{2,I}}$                             & 0.63          & 0.63          \\
$c_{\mathrm{1,II}}=c_{\mathrm{2,II}}$                           & 9.45          & 9.45                  \\
$l_{\mathrm{s}}$                                                & 0.1           & 0.1   \\
$c_{\mathrm{s}}$                                                & 7             & 7             \\
$K_{\mathrm{s}}$                                                & $5.012\times 10^{4}$ & $5.012\times 10^{4}$ \\        
\hline
\textit{IC radiation}                                  & \textit{A} & \textit{B}       \\
\hline
$l_{\mathrm{c}}$                                                & 0.1           & 0.1             \\
$c_{\mathrm{c}}$                                                & 7             & 7            \\
$K_{\mathrm{c}}$                                                & 5.012         & 5.012         \\
\hline
\end{tabular}
\end{table}

The time evolution of the low-energy branch of the spectrum, mainly resulting from synchrotron 
emission of the leptonic jet population, is common to both cases.
As shown in Figs.\ 3 and 4, the kinematic evolution of the helical jet described in Sect.\ 
\ref{LC} can well account for the SED changes corresponding to the radio-optical light curve 
periodic behaviour (Fig.\ 1): as the $\phi$ angle increases with the rotation and the first, 
strongly emitting, portion of the jet becomes more and more misaligned, the synchrotron peak 
slowly moves towards lower energies, finally plunging into the ``base level'' SED.

The high-energy part of the SED was studied by taking the historical X-ray spectral 
properties of the source into account. 
As previously mentioned, the shape of the X-ray spectrum is extremely variable, exhibiting 
alternatively hard and steep indices, and thus raising the controversial issue of the origin of 
the source X-ray emission. 
According to the value of $\alpha$, it could either be synchrotron radiation produced by 
high-energy electrons in the jet, or come from IC scattering of soft synchrotron photons (SSC) or 
external radiation (EC) off jet electrons, or be the result of a composition of synchrotron and IC 
effects.

We hence simulated the high-energy SED time evolution in two different cases.

In the first one (Fig.\ 3), we assumed that the whole X-ray jet emission comes from IC scattering
of softer synchrotron photons off jet energetic electrons [$\gamma_0 \approx 2\times 10^{5}$,
see Eq.\ (8)]. 

The X-ray flux variability and the constancy of the relevant spectral slope, exhibited by the 
modelled emission as the rotation goes on, reflect the radio SED evolution, while the 
$\gamma$-ray SED fluctuations correspond to the infrared--optical ones.
The parameter setting is shown in Col.\ \textit{A} of Table 1.

In the second case (Fig.\ 4), we supposed the existence of a synchrotron contribution to the soft
X-ray emission to justify the steepest spectra, like those observed by the Einstein Observatory 
in August 1980 and by ROSAT in July--August 1993; 
on the contrary, hard X-rays were still assumed to come from IC scattering of softer photons. 
The modelled jet is now characterized by a production of synchrotron radiation up to 
frequencies much higher than in the previous case ($\sim 10^{16}$ Hz) in the very first 
portion of the emitting jet.
The consequence is the appearance of a synchrotron ``bump'' in the SED of Fig.\ 4 in the soft X-ray 
band, well visible as long as the jet apex region is oriented at small angles 
($\phi \la 20^{\circ}$) with respect to the line of sight.
Moreover, the soft X-ray part of the SED becomes much more variable in both flux and slope, thus 
matching better the observed spectra.
The $\gamma$-ray modelled emission also exhibits remarkable new features.
It is the result of IC processing of a much wider range of frequencies and seems to better match 
observations with relatively low electron energies ($\gamma_{0}\approx 6 \times 10^{3}$).
The first SED peak occurs in the MeV band and corresponds to the infrared one, while the 
second peak lies in the GeV region and reflects the synchrotron X-ray bump, which thus would be 
the source of the seed photons for the EGRET detections. 
A complete description of the model parameters is given in Col.\ \textit{B} of Table 1.

\begin{figure*} [!hbtp]         \label{teta8mod}
%\sidecaption
\includegraphics[width=12cm,height=8.1cm]{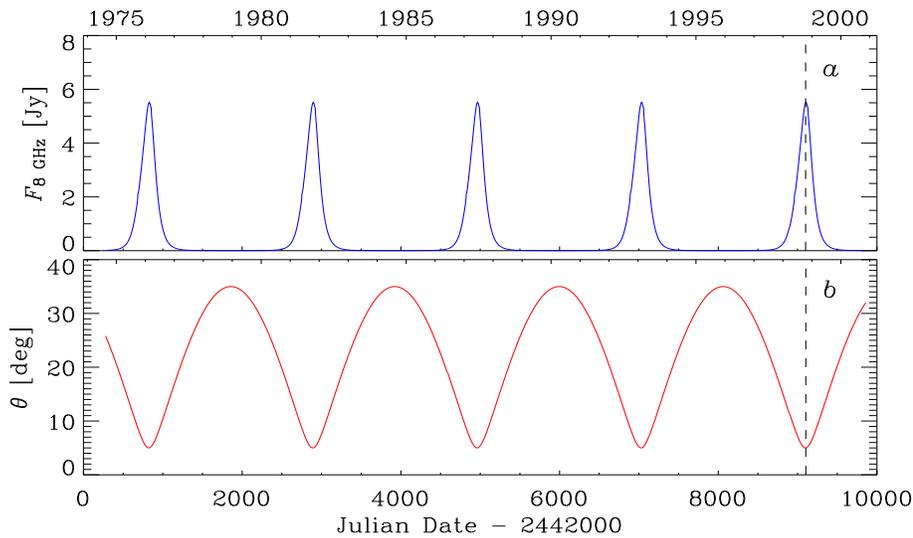}
\caption[]{
\textit{\ a}: Helical-jet modelling of the 8 GHz light curve.
\textit{\ b}: Time evolution, with the helix rotation, of the viewing angle $\theta$ of the jet 
            region emitting radiation observed at 8 GHz.
            Minima of the viewing angle $\theta$ correspond to maxima of 
            the flux, as marked by the vertical dashed line. 
            Only jet emission has been considered here.}
\end{figure*}

\begin{figure*} [!hbtp]         \label{teta8flares}
%\sidecaption
\includegraphics[width=12cm,height=7.1cm]{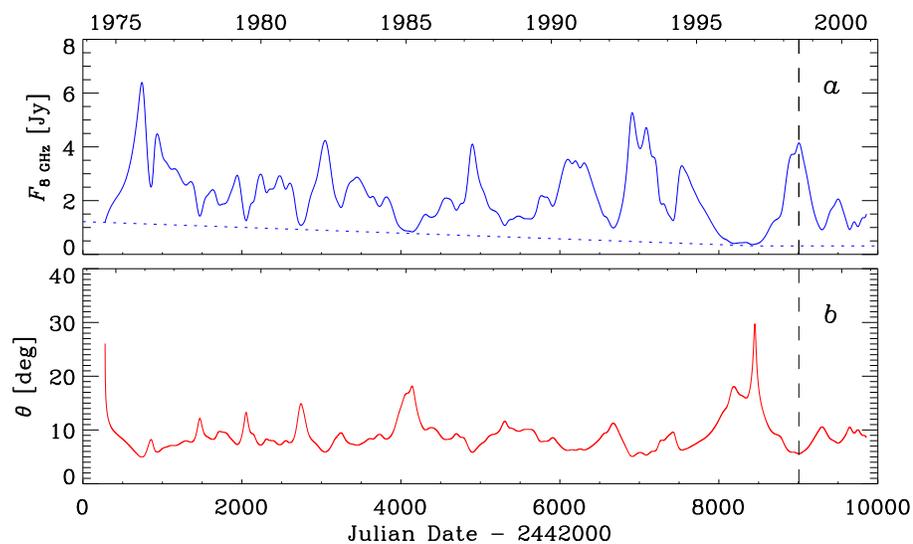}
\caption[]{
\textit{\ a}: Cubic spline interpolation through the binned (50 days/bin) 8 GHz light curve 
                (solid line) and ``base level'' flux (dotted line).
\textit{\ b}: Model time evolution of the viewing angle of the 8 GHz jet emitting region, 
                derived from the light curve of panel \textit{a} in the hypothesis that minor, 
                non-periodic outbursts are due to some distortions of the helical structure.}
\end{figure*}

\begin{figure*} [!hbtp]         \label{teta8noflares}
%\sidecaption
\includegraphics[width=12cm,height=7.1cm]{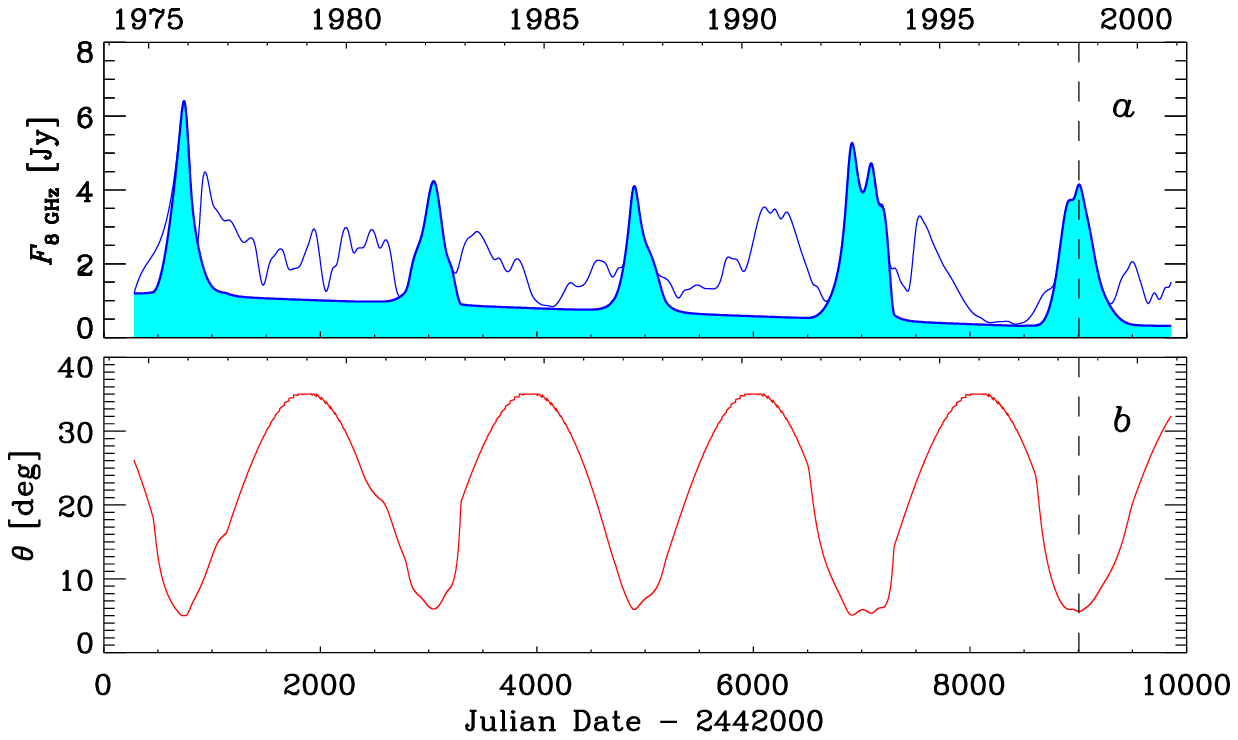}
\caption[]{
\textit{\ a}: The same as Fig.\ 5\textit{a}, with the periodic component highlighted by shading.
\textit{\ b}: Time evolution of the viewing angle corresponding to the shaded light
              curve of panel \textit{a}.}
\end{figure*}

\section{Modelling the non-periodic events: the 8 GHz light curve} \label{8GHz}

With the aim of investigating the variability of the source emission in deeper detail, we attempted 
a more accurate comparison between model and observations for the best-sampled
radio light curve at 8 GHz.

The model predicts that, as the helix rotates, the viewing angle of the 8 GHz jet emitting region 
varies periodically, as shown in Fig.\ 5\textit{b}, implying the light-curve periodic behaviour 
displayed in panel Fig.\ 5\textit{a}.

In order to make the comparison between model and data easier, we removed the short-term flaring 
from the observed curve of Fig.\ 1{\textit{c}} by deriving a binned light curve with a bin 
size of 50 days, and by fitting it with a cubic spline interpolation (Press et al. \cite{pre92}), 
as shown in Fig.\ 6{\textit{a}}.
The minor, non-periodic outbursts interspersed among the periodic major ones are clearly visible
in the curve. We can formulate two hypotheses about their origin:
(1) they are the effect of some geometric distortions of the helical path;
(2) they originate from other phenomena contributing to the source emission.

\begin{figure*} [!hbtp]         \label{LCpinching}
\centering
\includegraphics{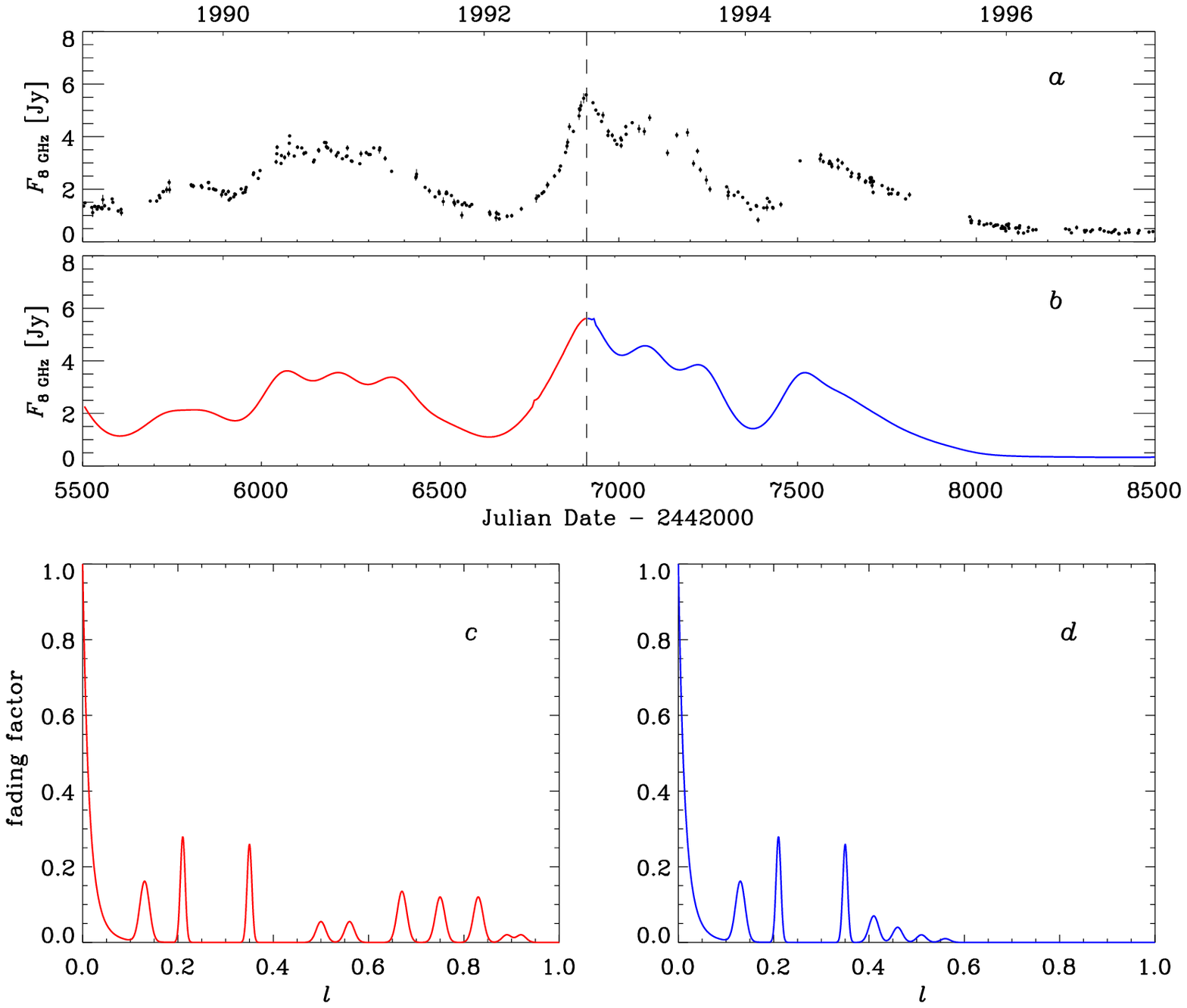}
\caption[]{\textit{a}: Observed 8 GHz light curve during an 8-year period centred on the 
1992--93 outburst.
\textit{b}: Modelled 8 GHz light curve, under the assumption that non-periodic outbursts
originate from the presence of enhanced-emission regions. 
\textit{c}: Emissivity fading factor before the 1992--93 periodic outburst.
\textit{d}: Emissivity fading factor after the 1992--93 outburst.}
\end{figure*}

By looking at the interpolation curve in Fig.\ 6\textit{a}, one can see that flux minima are 
less deep than simulated ones (Fig.\ 5\textit{a}), and that they seem to follow a linear decreasing 
trend down to the lowest state, as previously noticed for the radio light curves of panels 
\textit{b}, \textit{c}, \textit{d}, and \textit{g} of Fig.\ 1.
Hence we defined a ``base level'' for the source emission, represented by a straight line from the 
first datum of 1974 ($\rm JD=2442278.6469$) to the 1996 deepest minimum ($ \rm JD=2450454.7562$), 
and by a constant value in the remaining part of the curve (dotted line in Fig.\ 6\textit{a}). 
We then subtracted this trend from the spline, under the assumption that the resulting ``cleaned'' 
curve represents the jet contribution alone.
 
Following hypothesis (1), i.e.\  supposing the minor outbursts as due to distortions of the 
helical path, we derived the time evolution of the local viewing angle $\theta$ corresponding to the 
cleaned curve.
The difference between the result, displayed in Fig.\ 6\textit{b}, and the theoretical trend 
(Fig.\ 5\textit {b}) would represent the displacement of the jet path from a perfect helix in the 
jet region of interest.
The presence of the minor outbursts, missing in the modelled light curve, which keeps the mean flux 
high even when the base level is subtracted, is the reason why the average local viewing angle 
$\theta$ is low.

Under hypothesis (2), the minor outbursts come from other phenomena in the source. 
They might be explained as intrinsic variations of the emission, or as due to some more complex 
geometry of the jet structure implying other contributions to the flux.
By removing the non-periodic outburst component from the curve in Fig.\ 6\textit{a}, we obtained 
the shaded light curve of Fig.\ 7\textit{a}, and hence the time evolution of $\theta$ shown in 
Fig.\ 7\textit {b}, which looks very similar to that provided by the model, suggesting that 
the observed periodic behaviour can indeed be accounted for by a rotating jet structure 
fairly stable in time.

In this latter scenario, we tested the hypothesis that the non-periodic features of the light 
curve originate from some MHD instabilities occurring in the jet (e.g.\ pinching modes, 
which lead to the formation of jet regions characterized by higher-density magnetic field and 
stronger emissivity), or from other phenomena implying enhanced-emission regions (e.g.\ shock 
or magnetic reconnection events).

As the helix rotation goes on, these stronger emitting regions approach the line of sight, giving 
rise to outbursts in the observed light curve. 
The presence of enhanced-emission zones along the jet was simulated by modifying the 
emissivity fading factor represented in Fig.\ 2 (dashed, red line): Gaussian profiles were added to 
the curve to reproduce, as an example, the main light curve events observed at
8 GHz over an 8-year time interval centred on the 1992--93 periodic outburst.
The sequence of zones was supposed to be partially evolving during the selected period: 
the fading factor follows the trend displayed in Fig.\ 8\textit{c} before the outburst, and that in 
Fig.\ 8\textit{d} after the outburst.  
The corresponding modelled light curve, shown in Fig.\ 8\textit{b}, well reproduces the observed 
source behaviour, displayed in Fig.\ 8\textit{a}.

\section{Discussion and conclusions}

We found that the helical-jet model is able to describe the long-term behaviour of the multiwavelength 
emission of the BL Lacertae object AO 0235+16.
Both the periodic occurrence and the mean shape of the main radio and optical outbursts, as well as 
the corresponding remarkable variations of the broad-band SED, can be
explained in terms of the orientation change of an inhomogeneous, steadily-emitting, rotating helical jet. 
The radiation from the different-frequency emitting regions of the jet is affected by relativistic beaming, 
whose amount depends on the angle between the velocity vector of the emitting plasma and the line of sight, 
which changes along the helical path and also varies with time.
The assumption of flow instabilities (or other phenomena implying local
emission enhancement) in the jet provides a viable interpretation for the non-periodic
outbursts observed in the radio light curves.

The twisting of the jet in a helical structure and its rotation can originate from the orbital motion 
of the parent black hole in a BBHS, the main signature of which would indeed be the periodicity of the 
source light curves.

The periodicity of AO 0235+16 is about 2.9 years in the host galaxy rest reference frame, taking into 
account the $P=P_{\mathrm{obs}}/(1+z)$ relation with $z=0.94$.
According to Begelman et al. (\cite{beg80}), this period (assumed to be the orbital period) enables to 
estimate the mass of the primary black hole for any given value of the mass
ratio $M/m$ between the primary and secondary components:
$M \sim P_\mathrm{yr}^{8/5}(M/m)^{3/5}\,10^6\,M_{\sun}$.
For $M/m \sim$ 1--100 one can infer $M \sim 5 \times 10^6\,$--$\,9 \times 10^7\,M_{\sun}$; the
binary separation would be in the range $2\times 10^{-3}\,$--$\,5\times 10^{-3}\,\mathrm{pc}$.
If the jet is emitted by the primary, its orbital radius could vary from $5\times 10^{-5}\,$ up 
to $\,10^{-3}\,\mathrm{pc}$.

Several VLBI images of the source were produced over a wide range of observing radio wavelengths
in the past years. 
A large fraction of these maps shows no evidence of extended structures apart from the compact core, 
regardless of the different resolution (Gabuzda et al. \cite{gab92}; Gabuzda \& Cawthorne 
\cite{gabcaw96}); on the other hand, some of them reveal a faint jet north of the core
(Jones et al. \cite{jon84}; Chu et al. \cite {chu96}; Shen et al. \cite{she97}). 
The presence of a weak extension is confirmed by the sub-milliarcsecond maps obtained with the VSOP 
at 5 GHz (Frey et al. \cite{fre00}) and, more recently, with the VLBI at 43 GHz (Jorstad et al. 
\cite{jor01}), the latter also displaying a couple of components superluminally moving along bent 
trajectories.

Our model interprets the jet radio knots as the jet regions where the helical pattern presents the 
minimum viewing angle, with a maximization of the beaming effect.
According to the above orbital radius estimate and taking the pitch angle $\zeta$ and the viewing 
angle of the helix axis $\psi$ into account, the observed separation between radio knots should be 
less than 10 $\mu\mathrm{as}$, well below the resolution of the most detailed available maps.
All this under the assumption that the helix pitch does not vary when moving from the ``one-turn''
emitting region we considered, which is likely not true: the helix pitch could be smaller close 
to the black hole and growing outside, up to the observed sub-mas scales.

A huge observing effort is currently ongoing on this source, in order to closely follow its variability
behaviour around the time of the next predicted outburst (first half of 2004; Raiteri et al. \cite{rai01}):
optical and radio telescopes of the WEBT collaboration ({\tt http://www.to.astro.it/blazars/webt/}; 
Villata et al. \cite{vil00}, \cite{vil02}) are intensively monitoring it since summer 2003, together
with the Effelsberg 100 m radio telescope and VLBA.
Moreover, the ground-based observing effort will be intensified in 2004, during the optical/UV/X-ray
pointings of the source by the instruments onboard XMM-Newton, which will provide more details on 
the shape of the SED in the UV--X-ray band, possibly sheding light on the origin of X-rays in AO 0235+16.

\begin{acknowledgements}

      We wish to thank the anonimous referee for his/her useful comments, which have
      helped to clarify a few points of the paper,
      A. Ferrari for stimulating conversations,
      R. C.\ Hartman for useful suggestions on EGRET spectra,
      E.\ Trussoni for helpful information about the use of the XSPEC procedure,
      and A. Comastri for kindly providing us with details on the spectral
      analysis of the 1993 ROSAT data.

      This research has made use of:
        \begin{itemize}
        \item the NASA/IPAC Extragalactic Database (NED), which is operated by            
              the Jet Propulsion Laboratory, California Institute of Technology, 
              under contract with the National Aeronautics and Space Administration;

        \item the TARTARUS database, which is supported by Jane
              Turner and Kirpal Nandra under NASA grants NAG5-7385 and NAG5-7067;

        \item data from the University of Michigan Radio Astronomy Observatory,
              which is supported by the National Science Foundation and by funds from 
              the University of Michigan.      
        \end{itemize}

        This work was partly supported by the Italian Ministry for University and 
        Research (MURST) under grant Cofin 2001/028773, by the Italian Space Agency 
        (ASI) under contract CNR-ASI 1/R/27/02, and by the European Community's 
        Human Potential Programme under contract HPRN-CT-2003-00321, and it
        was part of the PhD thesis of L. O. at the University of Torino.

\end{acknowledgements}

\end{document}